\documentclass{PoS}

\title{Top Quark Physics At Hadron Colliders}

\ShortTitle{Top Quark Physics}

\author{\speaker{Fr\'ed\'eric D\'eliot}\\
        CEA-Saclay, Irfu/Spp\\
        E-mail: \email{frederic.deliot@cea.fr}}

\abstract{I summarize here the latest top quark physics results from the ATLAS, CDF, CMS and \dzero\ collaborations,
first discussing the top quark production and then some of the latest top quark property measurements.
These results are based on up to 5.4~\fb\ at the Tevatron and up to 1.10~\fb\ at the LHC.}

\FullConference{The 2011 Europhysics Conference on High Energy Physics, EPS-HEP 2011,\\
		July 21-27, 2011\\
		Grenoble, Rh\^one-Alpes, France}

\usepackage{rotating}
\RequirePackage{lineno}
\RequirePackage{cite}

\newcommand{\ttbar}  {\mbox{$t\bar{t}$}}

\newcommand{\ppbar}  {\mbox{$p\bar{p}$}}
\newcommand{\qqbar}  {\mbox{$q\bar{q}$}}

\newcommand{\fb}     {\mbox{fb$^{-1}$}}
\newcommand{\pb}     {\mbox{pb$^{-1}$}}
\newcommand{\dzero}  {\mbox{D0}}
\newcommand{\vtb}    {\mbox{$V_{tb}$}}
\newcommand{\mttbar} {\mbox{$M_{t\bar{t}}$}}
\newcommand{\pt}     {\mbox{$p_T$}}

\begin{document}

%\modulolinenumbers[1]
%\linenumbers

\section{Introduction}
\label{sec:intro}

Top quark physics is one of the main physics programs 
both at the Tevatron collider at Fermilab and 
at the Large Hadron Collider (LHC) at CERN.
The Tevatron is a proton-antiproton (\ppbar) collider with a center-of-mass energy of $\sqrt{s}=1.96$~TeV.
The LHC is colliding protons against protons with a current center-of-mass energy of $\sqrt{s}=7$~TeV.
Two experiments, CDF and \dzero, are located around the Tevatron, while the general purpose
detectors, ATLAS and CMS, are constructed around the LHC.
The Run~II of the Tevatron started in 2002 and terminated on September 30, 2011 with a delivered 
integrated luminosity of 11.9~\fb.
The LHC run at $\sqrt{s}=7$~TeV started in 2009. At the end of 2011, more than 5~\fb\ are expected 
to be delivered.

The top quark was discovered in 1995 by the CDF and D0 collaborations~\cite{topdiscovery}.
It is the heaviest elementary particle known today and has a coupling to the Higgs boson close to unity
which may indicate that it plays a special role in electroweak symmetry breaking.
Its lifetime is also shorter than the typical hadronization time thus it is the only quark that 
decays before hadronizing. That's why it offers an unique opportunity to study a bare quark.
For all these reasons, the top quark is a special quark and top quark physics is very relevant
to search for new physics.

Direct search for physics beyond the Standard Model (SM) has been performed in the top quark sector
by looking for specific new models that involve top quark signatures or for new particles 
that decay like top quarks. For instance searches for \ttbar\ resonance that
could be produced by the decay of a heavy Z' or searches for new couplings like 
flavor changing neutral currents were carried out. Direct searches in the top quark sector are described in~\cite{plenaries}.

It is also possible to search for new physics with top quarks in a model independent way looking
for deviations from the SM expectations. 
In that case it is necessary to precisely measure the top quark properties. New physics effects 
could be seen as new or anomalous couplings. For example, a heavy Z' exchange can be seen as a four fermion
coupling at low energy. An anomalous coupling between gluon and quarks would affect the production processes
$\qqbar \to \ttbar$ and $gg \to \ttbar$, while a four fermion operator would affect only the process
$\qqbar \to \ttbar$ and an anomalous three gluon coupling only the process:  $gg \to \ttbar$.
A new coupling between the $W$ boson and the quarks could affect both single top production and top 
quark decays. Hence different top quark observables can constrain different new physics effects 
and so it is useful to measure as many top quark properties as possible.

At hadron colliders the main top quark production occurs in pairs via the strong interaction by quark-antiquark
annihiliation or by gluon fusion. At the Tevatron, the dominant process is the quark-antiquark annihiliation
(85\% of the \ttbar\ production). The LHC is rather a gluon fusion machine since  $gg \to \ttbar$ represents 85\%
of the \ttbar\ production at $\sqrt{s}=7$~TeV.
The latest theoretical computations at an accuracy that approximates next-to-next-to-leading order (NNLO) in
perturbative theory gives: $\sigma(\ppbar \to \ttbar) = 7.46^{+0.48}_{-0.67}$~pb at the Tevatron and 
$\sigma(\ppbar \to \ttbar) = 164.6^{+11.4}_{-15.7}$~pb at the LHC at $\sqrt{s}=7$~TeV~\cite{ttbarxsectheo}
both for $m_t=172.5$~GeV.
Hence with 1~\fb\ at LHC, we expect around 4 times more \ttbar\ events than at the Tevatron with 5~\fb.

\begin{figure}[!htb]
\begin{center}
\includegraphics[height=25mm]{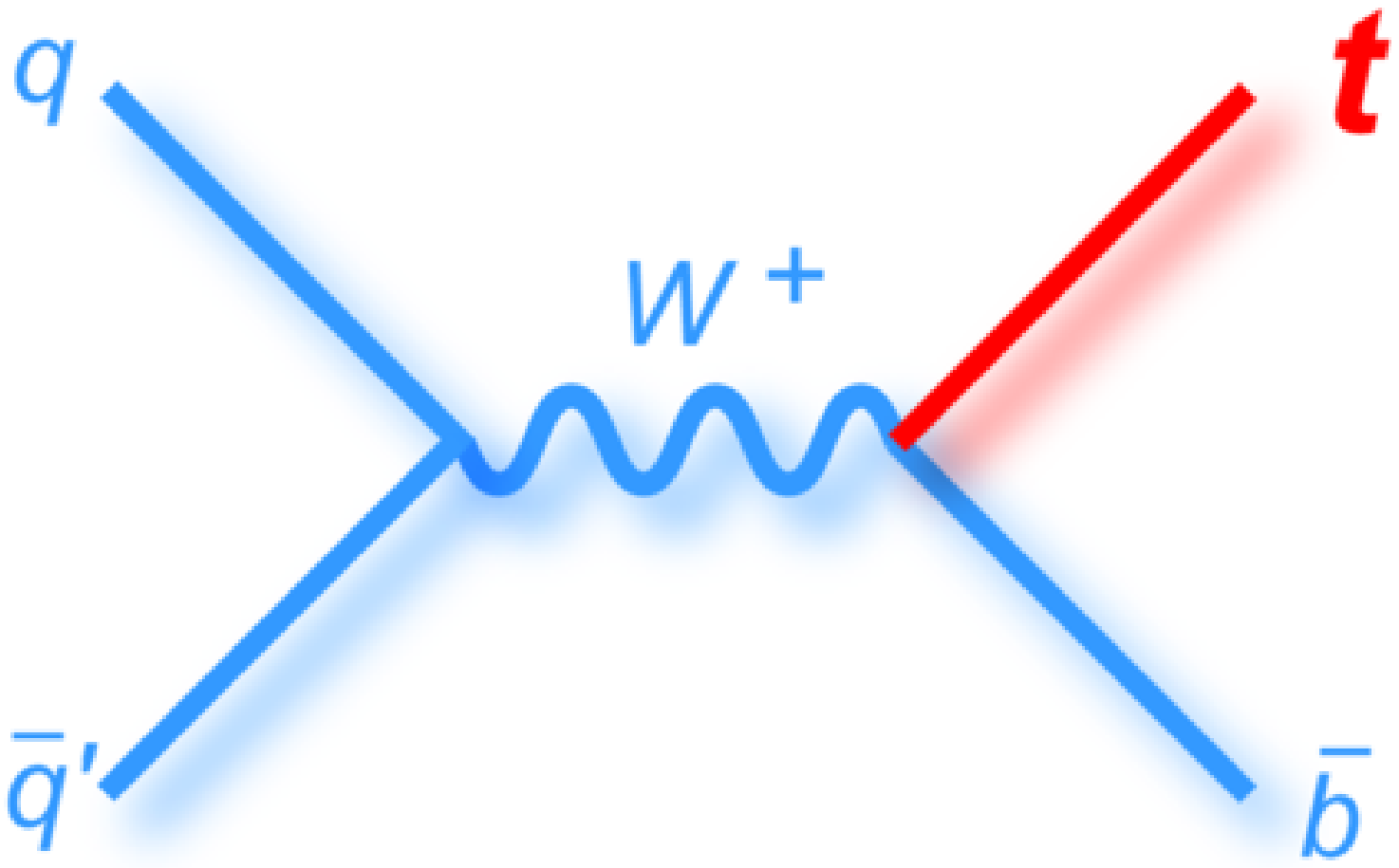}
\includegraphics[height=25mm]{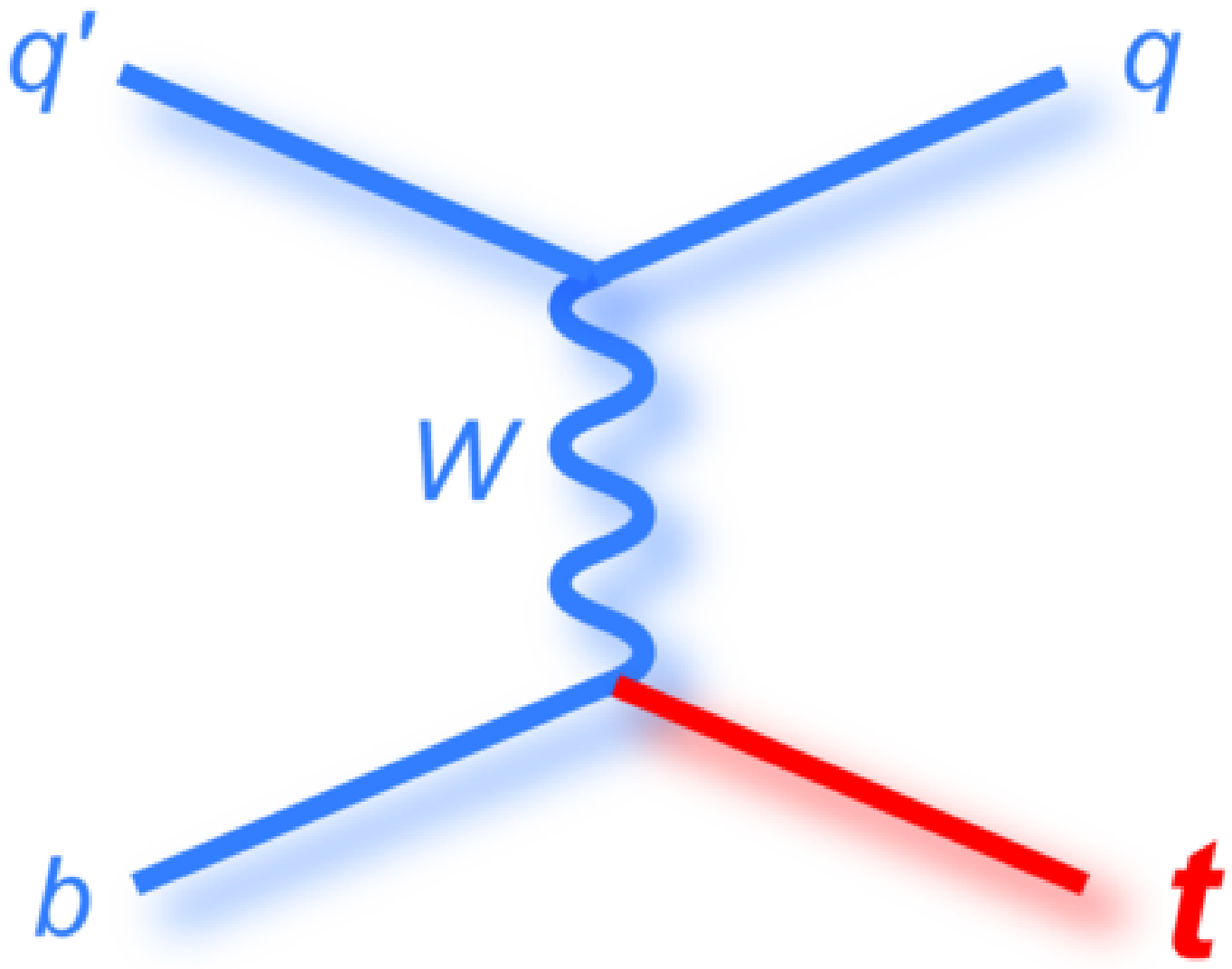}
\includegraphics[height=25mm]{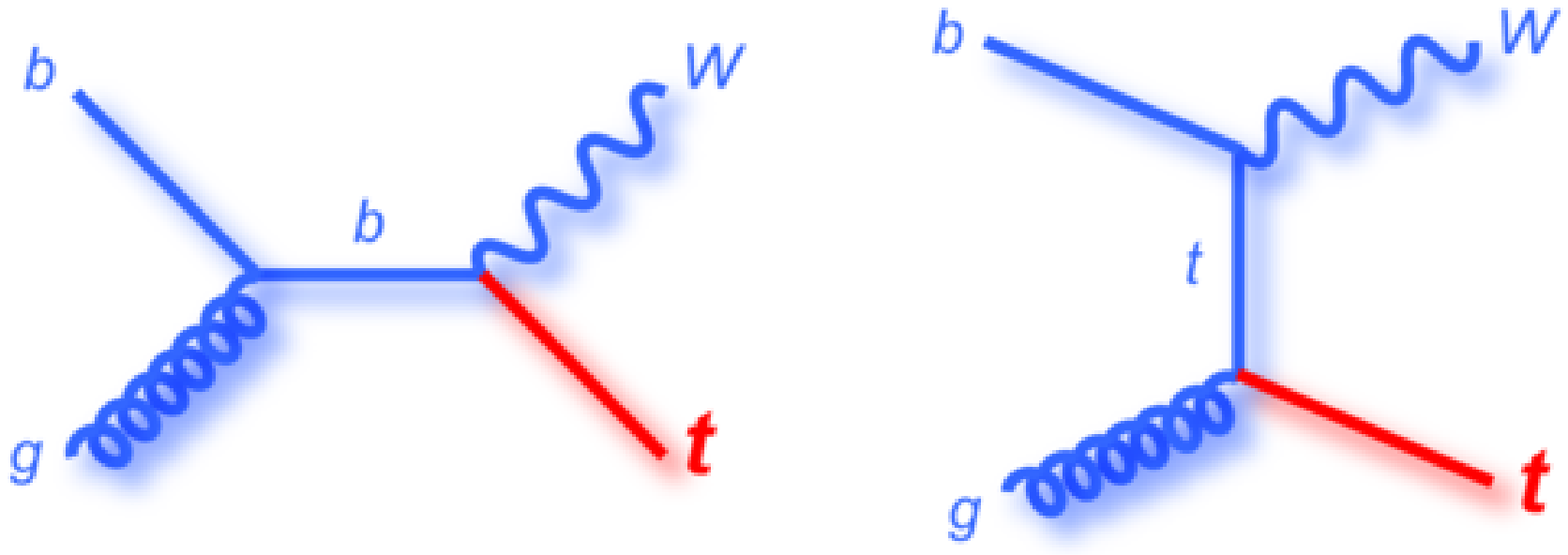}
\caption{Feynam diagrams for the production of single top at hadron colliders. 
From left to right: s-channel, t-channel and Wt-channel.}
\label{fig:stop}
\end{center}
\end{figure}

Top quarks can also be produced singly through the electroweak interaction. This production mode 
was discovered by CDF and \dzero\ in 2009~\cite{stopdiscovery}.
It allows to directly measure the \vtb\ element of the CKM matrix.
It is however challenging to measure since it has a rather small production cross section and
its background has a very similar signature to the single top signal. 
Single top production at hadron colliders can be separated into three channels.
The Feynam diagrams for the s-channel ($tb$), the t-channel ($tqb$) and the Wt-channel ($Wt$)
are shown in Figure~\ref{fig:stop}.

The latest theoretical computations are presented in Table~\ref{tab:sttheo}.
The t-channel is the dominant mode both at the Tevatron and the LHC,
while the s-channel is subdominant at the Tevatron and the Wt-channel
subdominant at the LHC.
It is interesting to measure these three processes separately since new physics 
could show up differently in the different channels. For instance a potential 
four quark coupling would affect only the s and t-channels while an anomalous
top-gluon coupling would alter only the Wt-channel. Anomalous Wtb coupling would
influence all three channels.
Due to its tiny cross section, measuring the Wt-channel is not possible at the Tevatron.
For the same reason, measuring the s-channel at LHC is challenging.

\begin{table}
\begin{center}
\begin{tabular}[t]{|l|ccc|}
\hline
in pb & $\sigma_{tb}$   & $\sigma_{tqb}$ & $\sigma_{tW}$ \\ \hline
\ppbar \ @ 1.96~TeV & $1.04 \pm 0.04$ & $2.26 \pm 0.12$ & $0.28 \pm 0.06$ \\
& \multicolumn{3}{c|}{\tiny PRD 74, 114012 (2006)} \\ \hline
pp @ 7~TeV         & $4.6 \pm 0.3$ & $64.6^{+3.3}_{-2.6}$ & $15.7 \pm 1.4$ \\
                   & {\tiny PRD81, 054028 (2010)}& {\tiny PRD83, 091503 (2011)} & {\tiny PRD82, 054018 (2010)} \\ \hline
\end{tabular}
\caption{Summary of the latest single top cross section computations for $m_t = 172.5$~GeV.}
\label{tab:sttheo}
\end{center}
\end{table}

Within the SM, the top quark decays almost 100\% of the time into a W boson and a $b$ quark.
This branching ratio can be modified by the presence of new physics.
Top pair signatures are then classified according to the decays of the W bosons.
If both W bosons from the top quarks decay hadronically, the channel is called the alljets
final state. It has a large rate but is also contaminated by a large background from multijet events
that is estimated from data.
On the contrary, the dilepton channel occurs when both W bosons decay leptonically into a muon or an
electron. It has a small branching ratio but also a small background contamination. The main 
background in this channel comes from Drell-Yan production (with fake missing transverse energy in the 
dielectron or dimuon channels).
The golden channel is the lepton+jets channel when one W boson from the top quark decays leptonically
and the other one hadronically. It has a good rate with reasonable background from W+jets production and multijet events.
In this channel the W+jets background normalization is usually estimated from data while its shape is taken from MC.
As top quark decays always produce $b$ quarks, $b$ quark identification is often used to enhance the purity in 
the selected data samples.

In the following, I will describe first the latest results on top quark production then I will present the measurements
of the top quark properties. References for the preliminary results presented here can be found on the collaboration
public web pages~\cite{cdf,dzero,atlas,cms}.

\section{Top Quark Production}
\label{sec:prod}

Studying top quark production mainly consists of measuring the top quark production cross sections. 
These measurements also allow to handle well known data samples that can be further used to scrutinize the
other top quark properties.

\subsection{\ttbar\ Production Cross Section}
The most precise results on the \ttbar\ cross section are measured in the lepton+jets channel.
Such measurements could be based on purely topological information or could be using identification
of jets from $b$-quarks ($b$-tagging requirements).
As the main background in this channel comes from W+jets events, the rate of which is difficult to predict theoretically, 
the normalization of this background is usually fit together with the number of \ttbar\ events.
It is also valuable especially in this channel to use the data to constrain systematic uncertainties  
in order to reduce them.

The most precise measurements of the \ttbar\ cross section in this channel are summarized in Table~\ref{tab:ljxs}
both at the Tevatron and at the LHC.
All these measurements are limited by systematic uncertainties where the largest sources come from the uncertainty
on jet energy scale (JES) calibration, jet identification, $b$-tagging requirement and from the fraction of W events produced
in association with heavy flavor quarks.

\begin{table}
\begin{center}
\begin{tabular}[!htb]{|l|ll|}
\hline
CDF (4.6 \fb, {\small PRL 105, 012001 (2010)}) & $\sigma(\ppbar \to \ttbar) =$ & $7.70 \pm 0.52 \ {\rm (stat+syst+theory) \ pb}$ \\[6pt]
D0  (5.6 \fb, {\small PRD 84, 012008 (2011)}) & $\sigma(\ppbar \to \ttbar) =$ & $7.78^{+0.77}_{-0.64} \ {\rm (stat+syst+lumi) \ pb}$ \\[6pt]
\hline \\[-12pt]
Atlas (35 \pb) & $\sigma(pp \to \ttbar) =$ & $186 \pm 10 {\rm (stat)} ^{+21}_{-20} {\rm (syst)} \pm 6 {\rm (lumi) \ pb}$ \\[6pt]
CMS   (36 \pb, {\small EPJ. C71, 1721 (2011)}) & $\sigma(pp \to \ttbar) =$ & $150 \pm 9 {\rm (stat)} \pm 17 {\rm (syst)} \pm 6 {\rm (lumi) \ pb}$ \\
\hline
\end{tabular}
\caption{Summary of the latest \ttbar\ cross section measurements in the lepton+jets channel for $m_t = 172.5$~GeV.}
\label{tab:ljxs}
\end{center}
\end{table}

It is also interesting to measure the \ttbar\ production rate in other channels since new physics
could affect the top quark decay channels differently. It is then important to perform the measurements in different
signal/background environment and to see if the measurements in all decay channels agree with each other.
Measurements are now evaluable in almost all possible top quark decay channels. In particular, ATLAS has provided a new measurement in the 
dilepton channel using 0.70~\fb, while CMS measured for the first time the \ttbar\ cross section in the alljets and $\mu\tau$ 
decay channels using 1.10~\fb. The measurements in all the different channels are in good agreement.

Cross section measurements can also be used as a tool to study other properties. For instance, \dzero\ recently 
fit, together with the \ttbar\ cross section, the branching fraction ratio $R = \frac{B(t \to Wb)}{B(t \to Wq)}$, 
where $q$ represents any type of quarks. This ratio which is predicted to be 1 in the SM can be expressed in term of 
elements of the CKM matrix and can then be used to measure $V_{tb}$ assuming the unitarity of the CKM matrix. 
Using 5.4~\fb\ in both the lepton+jets and dilepton channel, \dzero\ 
measures: $|V_{tb}|=0.95 \pm 0.02$~\cite{d0Rb}.

\begin{figure}[!htb]
\begin{center}
\includegraphics[height=70mm]{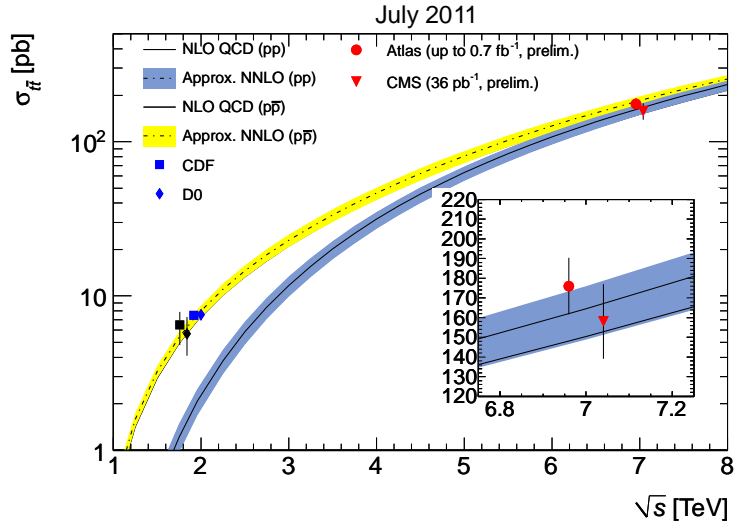}
\caption{Summary of the latest \ttbar\ cross section measurements at the Tevatron ad LHC as a function of the center-of-mass energy.}
\label{fig:allxs}
\end{center}
\end{figure}

The latest cross section measurements at the Tevatron and LHC are summarized in Table~\ref{tab:allxs} and in Figure~\ref{fig:allxs}.
The precision of the measurements is around 6.5\% at the Tevatron and as low as 8\% at the LHC.
The results agree with the Quantum Chromodynamics (QCD) predictions. 
In the future, precise differential measurements will also be carried out.

\begin{table}
\begin{center}
\begin{tabular}[!htb]{|l|ll|}
\hline
CDF (up to 4.6 \fb) & $\sigma(\ppbar \to \ttbar) =$ & $7.5 \pm 0.31 {\rm (stat)} \pm 0.34 {\rm (syst)} \pm 0.15 {\rm (theory) \ pb}$ \\[6pt]
D0  (5.6 \fb, {\small PLB 704, 403, (2011)}) & $\sigma(\ppbar \to \ttbar) =$ & $7.56^{+0.63}_{-0.56} \ {\rm (stat+syst+lumi) \ pb}$ \\[6pt]
\hline \\[-12pt]
Atlas (up to 0.7 \fb) & $\sigma(pp \to \ttbar) =$ & $176 \pm 5 {\rm (stat)} ^{+13}_{-10} {\rm (syst)} \pm 7 {\rm (lumi) \ pb}$ \\[6pt]
CMS   (36 \pb) & $\sigma(pp \to \ttbar) =$ & $158 \pm 10 {\rm (uncor.)} \pm 15 {\rm (cor.)} \pm 6 {\rm (lumi) \ pb}$ \\
\hline
\end{tabular}
\caption{Summary of the latest combined \ttbar\ cross section measurements for $m_t = 172.5$~GeV.}
\label{tab:allxs}
\end{center}
\end{table}

\subsection{Single Top Production Cross Section}

As discussed above, the dominant mechanism for electroweak top quark production is the t-channel production where the top quark
is produced in association with a light and a $b$~quark.
To measure the t-channel cross section, the analysis strategy consists of isolating this particular topology from the other
single top processes and from the backgrounds (mainly W+jets and \ttbar). 
Due to the small production rate and the large background, multivariate methods like neural networks (NN) or boosted decision trees (BDT)
have to be deployed at the Tevatron. At the LHC both cut-based or multivariate approaches are used.

The latest \dzero\ measurement~\cite{d0tchannel} uses a combination of three multivariate discriminants to extract 
the t-channel single top cross section and reaches an observed significance of $5.5~\sigma$ for this measurement.
ATLAS uses both a cut based and a NN which allow also to observe the signal at the level of $7.6~\sigma$.
CMS uses a 2 dimension fit of the cosine of the angle between the lepton and the light jet and the pseudorapidity of the light 
jet~\cite{cmstchannel}.
CMS also measures the t-channel cross section using a BDT~\cite{cmstchannel}. 
Table~\ref{tab:tchannel} summarizes the latest t-channel cross section measurements at the Tevatron and the LHC.
\begin{table}
\begin{center}
\begin{tabular}[!htb]{|l|ll|}
\hline
CDF (3.2 \fb)                           & $0.8 \pm 0.4$ & \\[3pt]
D0  (5.4 \fb, {\small PLB 705, 313 (2011)}) & $2.90 \pm 0.59$ & $5.5 \sigma$ \\[3pt] \hline
CMS (36 \pb,  {\small PRL 107, 091802 (2011)}) & $83.6 \pm 29.8 {\rm (stat+syst)} \pm 3.3 {\rm (lumi)}$ & $3.7 \sigma$ \\[3pt]
Atlas (0.7 \fb)         & $90^{+32}_{-22}$ & $7.6 \sigma$ \\ 
\hline
\end{tabular}
\caption{Summary of the t-channel cross section measurements for $m_t = 172.5$~GeV (in pb).}
\label{tab:tchannel}
\end{center}
\end{table}

Measurements of the other single top cross sections have been also performed.
\dzero\ performed a measurement of the inclusive cross section using both t and s-channels as signal using 5.4 \fb\ of data,
leading to: $3.43^{+0.73}_{-0.74}$ pb for $m_t = 172.5$~GeV~\cite{d0stop}. This allows to extract a limit on $V_{tb}$ of:
$|V_{tb}|>0.79$ at 95~\% confidence level (CL). More statistics is needed to be sensitive to the s-channel.
ATLAS also performed a search for the Wt-channel in the dilepton channel using a cut based analysis.
In this channel, the main background comes from \ttbar\ production. A limit at 95~\% CL of 
$\sigma_{Wt}<39.1$~pb is achieved.

\section{Top Quark Properties}
\label{sec:properties}

Many top quark properties have been measured already. We will focus here only on some of the most recent developments.

\subsection{Top Quark Mass}
It is important to precisely measure the top quark mass since it is a free parameter of the SM and because
together with the mass of the W boson it allows to predict the mass of the yet unobserved Higgs boson.
If the Higgs boson is discovered, it would allow to test the consistency of the SM.

There are mainly three different methods to measure directly the top quark mass. First the template method
which relies on the comparison of an observable in data (often the reconstructed top quark mass itself) with 
the prediction for this observable from MC samples generated at different masses. The main advantage of this 
method is its simplicity. The second method called the matrix element method leads to the most precise determination
of the top quark mass. It consists of building an event by event probability based on the Leading Order (LO) \ttbar\ matrix 
element using the full kinematics of the event. The third method called the ideogram method can be seen as an approximation
of the matrix element method. It uses an event likelihood computed as a convolution of a Gaussian resolution function 
with a Breit-Wigner for the signal.
Independently of the method, for the channels with a least one W boson decaying hadronically, it is possible to 
calibrate the JES constraining the invariant mass of the two light jets from the W decays to the 
world average W mass. This allows to greatly decrease the input of the uncertainty from JES. 
In order to correct for any potential biases due to the method assumptions it is mandatory to calibrate the measurements.

The results for some of the latest measurements of the top quark mass are given below.
CDF measured the mass requiring missing transverse energy in associated with jets. It allows
to recover acceptance for misidentified leptons from the lepton+jets channel. 
It extracts the signal using a NN. The top quark mass is determined using a template method with three jet invariant masses as 
observable for the top quark mass and dijet invariant masses to calibrate in-situ the JES.
This method leads to $m_{\rm top} = 172.3 \pm 2.4 \ (\rm stat+JES) \pm 1.0 \ (\rm syst)$~GeV.
\dzero\ measured the top quark mass using the matrix element method in the lepton+jets channel.
This updated measurement benefits from a new flavor-dependent jet response correction that allows 
to additionnally reduce the JES uncertainty. Combining the 2.6 \fb\ new result with the previous published 
1~\fb\ measurement leads to: $m_{\rm top} = 174.94 \pm 0.83 \ (\rm stat) \pm 0.78 \ (\rm JES) \pm 0.96 \ (\rm syst)$~GeV~\cite{d0ljmass}. 
This measurement is limited by systematic uncertainties from signal modeling and from the residual JES uncertainty.
CMS used an ideogram method in the lepton+jets channel using 36 \pb\ and extracted a top quark mass of: 
$m_{\rm top} = 173.1 \pm 2.1 \ (\rm stat) ^{+2.8}_{-2.5} \ (\rm syst)$~GeV. This result is also limited by systematic uncertainties 
mainly from JES.

\begin{figure}[!htb]
\begin{center}
\includegraphics[width=100mm]{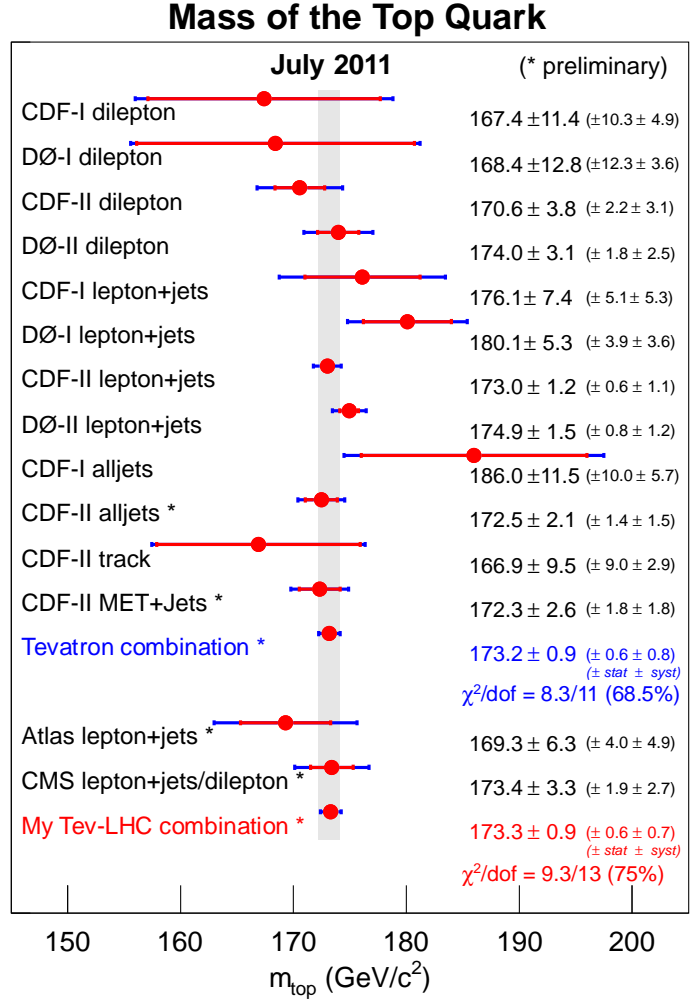}
\caption{Summary of the latest Tevatron top quark mass measurements and their combined value~\cite{tevmasscombi}.
My attempt to add the LHC measurements in the combination is also shown.}
\label{fig:tevmasscombi}
\end{center}
\end{figure}

CDF and \dzero\ have updated the combination of their top quark mass measurements in all the different top quark decay channels~\cite{tevmasscombi}.
All channels lead to consistent results and for the first time, the combined value has an uncertainty below 1~GeV as can be seen 
in Figure~\ref{fig:tevmasscombi}. Work is still in progress to further reduce the systematic uncertainties.
I made an attempt to add the LHC measurements in the combination taking correlations into account. 
The result is also shown in  Figure~\ref{fig:tevmasscombi}.

Using this new Tevatron top quark mass combination, electroweak fits constraint the Higgs boson mass to be:
$m_H < 161$~GeV at 95~\% CL with a most probable value of $m_H = 92 ^{+34}_{-26}$~GeV~\cite{ewfit}.

\subsection{W Boson Helicity in Top Decays}
Measuring the helicity of the W boson in top quark decays enables to test the SM at the electroweak scale. 
New physics could affect the helicity through the coupling of the W boson to the top and bottom quarks.
The SM predicts that the W boson can not be right-handed.

\begin{figure}[!htb]
\begin{center}
\includegraphics[width=100mm]{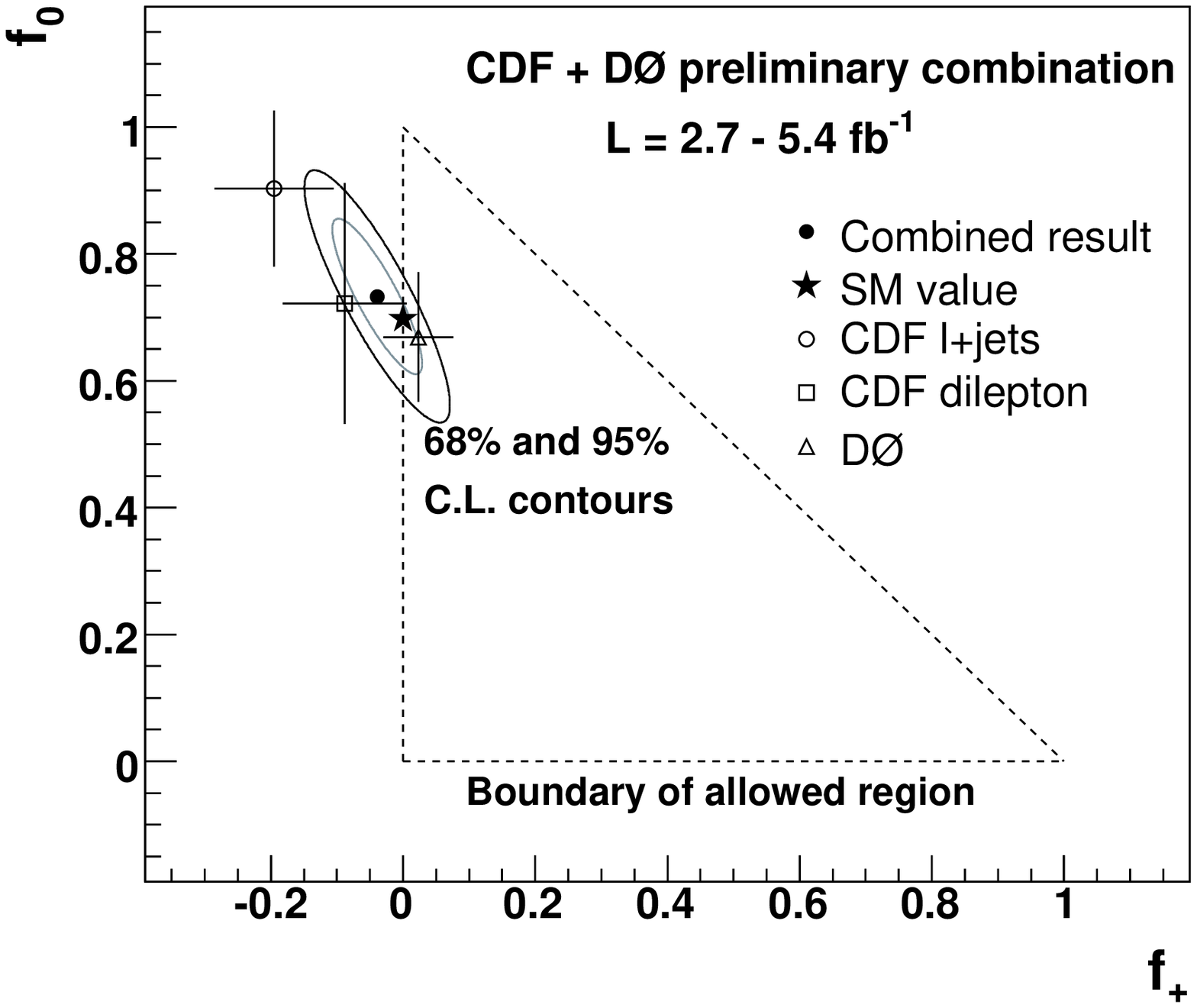}
\caption{Summary of the latest Tevatron W helicity measurements and their combination.}
\label{fig:tevwhel}
\end{center}
\end{figure}

The measurement is performed either using a template fit or using a matrix element method.
For the template method, the often chosen observable is the cosine of the angle between
the lepton from the W boson decay and the top quark direction in the W boson rest frame ($\cos \theta^*$).
Combining the latest Tevatron measurements in the dilepton and lepton+jets channels taken into account correlations leads to:
$f_0 = 0.732 \pm 0.063 \ (\rm stat) \pm 0.052 \ (\rm syst)$ and $f_+ = -0.039 \pm 0.034 \ (\rm stat) \pm 0.030 \ (\rm syst)$
when both the left-handed ($f_0$) and right-handed ($f_+$) W fractions are allowed to vary.
This measurement agrees with the SM predictions of $f_0 = 0.70$ and $f_+ = 0$.
The input measurements as well as the combined result are shown in Figure~\ref{fig:tevwhel}.

Altas also measured the left-handed and longitudinal ($f_-$) W fractions in the lepton+jets channel using
35 \pb and finds: $f_0 = 0.59 \pm 0.10 \ (\rm stat) \pm 0.07 \ (\rm syst)$ when $f_+$ and $f_-$ are fixed to 
their SM values and $f_- = 0.41 \pm 0.10 \ (\rm stat) \pm 0.07 \ (\rm syst)$ when $f_0$ and $f_+$ are fixed.

\subsection{Top Pair Spin Correlations}
Even if at the Tevatron and LHC, the top quarks are produced unpolarized, in the SM, the spin of the top 
and of the antitop quarks are correlated. Since the top quark decays before hadronizing, this correlation is
preserved in the decay products. Hence the quark is a unique tool to study this property.
These correlations can be affected by new physics. Measuring the correlations allow also to verify that
the top quark is a spin 1/2 particle. In quark-antiquark annihilation, the top pair is produced in a $^3 S_1$ state
while it is produced in a $^1 S_0$ state via gluon fusion. That's why this correlation is different at the Tevatron
and at the LHC.  

It is possible to measure the top spin correlation using a template or a matrix element method.
As the differential \ttbar\ cross section $\sigma$ can be written: 
$$\frac{1}{\sigma} \frac{d^2 \sigma}{d \cos \theta_1 d \cos \theta_2} = \frac{1}{4} (1 - C \cos \theta_1 \cos \theta_2),$$ 
the  template method relies on a fit of the $\cos \theta_1 \cos \theta_2$ distribution where
$\theta$ is the angle between the down-type fermion with respect to the chosen spin basis in the top or antitop
quark rest frame and allows to measure the correlation strength $C$.
CDF measured the spin correlation when choosing the helicity basis as quantization axis using 4.3~\fb\ in the lepton+jets channel 
and found $C= 0.60 \pm 0.50 \ (\rm stat) \pm 0.16 \ (\rm syst)$~\cite{cdfspin}. The SM predicts $C = 0.40$.
\dzero\ has developped a new method by measuring the fraction of events with spin correlation using a template fit of the variable
$$R = \frac{P(H=c)}{P(H=u) + P(H=c)}$$ where $P(H=c)$ is the probability that the signal has spin correlation and 
$P(H=u)$ the probability that the signal has no spin correlation. The probabilities are computed using the LO \ttbar\ 
matrix element with or without spin correlation. Using 5.4 \fb\ of dilepton events, \dzero\ measured 
$f = 0.74^{+0.40}_{-0.41} \ (\rm stat+syst)$ which can be translated using the NLO prediction of $C$ in the SM 
to $C = 0.57 \pm 0.31 \ (\rm stat+syst)$~\cite{d0spin}.
Even if these measurements are still statistically limited, their sensitivity to exclude the case of no correlation
is now close to 3~$\sigma$.

\subsection{Top-Antitop Charge Asymmetry}
At Next-to-Leading Order (NLO), perturbative QCD predicts an asymmetry for \ttbar\ events produced
via quark-antiquark annihilation. Indeed, top quarks are predicted to be emitted preferably in the direction of the 
incoming quarks. The exchange of new particles like a Z' or an axigluon could modify it.
This asymmetry comes from the interference of the $\qqbar \to \ttbar$ tree and box diagrams and leads to a positive asymmetry
and the interference between the ISR and FSR $\qqbar \to \ttbar g$ diagrams which produces a smaller negative asymmetry. 
At the Tevatron which is a proton-antiproton collider, this asymmetry translates into a forward-backward asymmetry.
It can be measured with the observable: $$A^{t \bar{t}} = \frac{N(\Delta y > 0) - N(\Delta y < 0)}{N(\Delta y > 0) + N(\Delta y < 0)}$$  
where $\Delta y = y_t - y_{\bar{t}}$ is the difference between the top and antitop quark rapidities.
At the LHC which is a proton-proton collider, since the antitop quarks are coming from the sea, they carry on average less momentum than the 
incoming quarks. As the produced top quarks are emitted preferably in the direction of the incoming quarks due to the boost, they 
appear more central than antitop quarks. Hence at the LHC, the asymmetry can be observed as a central/forward-backward asymmetry.
A possible asymmetry observable is then:
$$A_C = \frac{N(\Delta |y|>0) - N(\Delta |y|<0)}{N(\Delta |y|>0) + N(\Delta |y|<0)}$$ where $\Delta |y| = |y_t| - |y_{\bar{t}}|$.
However, because at LHC the main \ttbar\ production process occurs via gluon fusion, this asymmetry is small.

CDF measured $A^{t \bar{t}}$ in the lepton+jets and dilepton channels.
After correcting for acceptance and reconstruction effects, the measurement from CDF in the lepton+jets channel shows a 
result 3.4~$\sigma$ higher than the SM prediction for $\mttbar > 450$~GeV~\cite{cdfasymmetry}. 
CDF measurements are summarized in Table~\ref{tab:cdfasymmetry}. 
\begin{table}
\begin{center}
\begin{tabular}[t]{|l|ccc|}
\hline
$A^{t\bar{t}}$      &  l+jets & l+jets ($\mttbar \ge 450$~GeV) & dilepton \\ \hline
unfolded data     & $0.158 \pm 0.074$  & $0.475 \pm 0.114$    & $0.42 \pm 0.16$ \\
SM prediction (MCFM~\cite{mcfm}) & $0.058 \pm 0.009$  & $0.088 \pm 0.013$    & $0.06 \pm 0.01$ \\
\hline
\end{tabular}
\caption{Summary of CDF top charge asymmetry measurements with the SM predictions.}
\label{tab:cdfasymmetry}
\end{center}
\end{table}

\dzero\ also measured the asymmetry using 5.4~\fb\ in the lepton+jets channel. After correction for detector effects, 
\dzero\ found $A^{t \bar{t}} = 0.196 \pm 0.065$~\cite{d0asymmetry} with agrees at the level of 2.4~$\sigma$ with the 
prediction from MC@NLO~\cite{mcatnlo}. No significant discrepancy was observed with the predictions at large $\mttbar$
at the reconstruction level.
Using lepton+jets events, \dzero\ also measured the leptonic asymmetry: 
$$A^l_{FB} = \frac{N(q_l y_l>0)- N(q_l y_l<0)}{N(q_l y_l>0)+ N(q_l y_l<0)}$$ where $q_l$ and $y_l$ are the lepton charge and rapidity.
After correction for detector effects, it gives: $A^l_{FB} = 0.152 \pm 0.04$~\cite{d0asymmetry} while the MC@NLO prediction 
leads to $0.021 \pm 0.001$.
This corresponds to a difference of more than 3~$\sigma$. \dzero\ also noticed that the measured asymmetry depends significantly 
on the modeling of the \pt\ of the \ttbar\ system which is not perfectly described by MC@NLO at \dzero~\cite{d0asymmetry}. 

At the LHC, ATLAS and CMS used slightly different observables to measure: 
$$A_C = \frac{N(\Delta>0) - N(\Delta<0)}{N(\Delta>0) + N(\Delta<0)}.$$ ATLAS utilized: $\Delta = |y_t| - |y_{\bar{t}}|$
while CMS used both: $\Delta = |\eta_t| - |\eta_{\bar{t}}|$ (where $\eta_{t/\bar{t}}$ are the pseudorapidity of the top and antitop quarks)
and $\Delta = (y_t-y_{\bar{t}})(y_t+y_{\bar{t}})$. The LHC results are summarized in Table~\ref{tab:lhcasymmetry}. No significant
discrepancies from the SM are observed so far. CMS asymmetry distribution as function \mttbar\ at the reconstruction level does 
not show any excess for large \mttbar.
\begin{table}
\begin{center}
\begin{tabular}[t]{|l|cc|}
\hline
                         &  unfolded data & SM prediction \\ \hline
Altas: $A^y_C$ (0.7 \fb)  & $-0.024 \pm 0.016 \ {\rm (stat)} \pm 0.023 \ {\rm (syst)}$ & 0.006 (MC@NLO) \\
CMS: $A^\eta_C$ (1.1 \fb) & $-0.016 \pm 0.030 \ {\rm (stat)}^{+0.010}_{-0.019} \ {\rm (syst)}$ & $0.0130 \pm 0.001$~\cite{rodrigo} \\
\hline
\end{tabular}
\caption{Summary of the Atlas and CMS top charge asymmetry measurements with the corresponding SM predictions.}
\label{tab:lhcasymmetry}
\end{center}
\end{table}

\section{Summary and conclusion}
\label{sec:summary}

Numerous top quark properties have been already measured allowing to better understand this unique quark and 
to test the SM at the electroweak scale. These measurements are summarized in Table~\ref{tab:summary}. 
With the exception of the puzzling top charge asymmetry, no deviations from the SM predictions have been observed. 
However only half of the Tevatron dataset has been analyzed so far.
With limited statistics, the LHC experiments have already delivered impressive top quark measurements
but a lot more precise measurements are expected when analyzing several \fb.
We are then looking forward for hopefully exciting discoveries in the top quark sector in the future.

\begin{center}

\begin{table}
\begin{center}
\begin{tabular}[!htb]{|l|ll|c|c|}
\hline
{\bf Property} & \multicolumn{2}{c|}{\bf Measurement} & {\bf SM } & {\bf {\cal{L}} (\fb)} \\[3pt] \hline
{\boldmath $\sigma_{\ttbar}$}
			& $\ppbar \to \ttbar $ & CDF: $7.5 \pm 0.31 {\rm (stat)} \pm 0.34 {\rm (syst)} \pm 0.15 {\rm (th)}$ pb & $7.46^{+0.48}_{-0.67}$ pb & up to 4.6 \\[3pt]
{\footnotesize (for $M_t=172.5$~GeV)}  & 		       & D0:  $7.56^{+0.63}_{-0.56} \ {\rm (stat+syst+lumi)}$ pb & 		 & 5.6 \\[3pt]
  & $pp \to \ttbar$      & Atlas: $180 \pm 9 {\rm (stat)} \pm 15 {\rm (syst)} \pm 6 {\rm (lumi)}$ pb & $164.6^{+11.4}_{-15.7}$~pb & up to 0.7 \\[3pt]
  & 		       & CMS:  $158 \pm 10 {\rm (uncor.)} \pm 15 {\rm (cor.)} \pm 6 {\rm (lumi)}$ pb & & 0.036 \\[3pt] \hline

 {\boldmath $\sigma_{\rm tbq}$}
	& $\ppbar \to \ttbar $ & CDF: $0.8 \pm 0.4$ pb ($M_t=175$~GeV) & $2.26 \pm 0.12$ pb & 3.2 \\[3pt]
{\footnotesize (for $M_t=172.5$~GeV)} & 		       & D0:  $2.90 \pm 0.59$ pb &  & 5.4 \\[3pt]
 & $pp \to \ttbar$      & Atlas: $90^{+32}_{-22}$ pb & $64.6^{+3.3}_{-2.6}$ pb & 0.7 \\[3pt]
 &			& CMS: $83.6 \pm 29.8 {\rm (stat+syst)} \pm 3.3 {\rm (lumi)}$ pb & & 0.035 \\[3pt] \hline
 {\boldmath $\sigma_{\rm tb}$}
			& $\ppbar \to \ttbar $ & CDF: $1.8^{+0.7}_{-0.5}$ pb ($M_t=175$~GeV) & $1.04 \pm 0.04$ pb & 3.2 \\[3pt]
{\footnotesize (for $M_t=172.5$~GeV)} &		       & D0: $0.68^{+0.38}_{-0.35}$ pb & & 5.4 \\[3pt] \hline
 {\boldmath $\sigma_{\rm Wt}$}
			& $pp \to \ttbar$      & Atlas: $< 39.1$ pb & $15.7 \pm 1.4$ pb & 0.7 \\[3pt]
{\footnotesize (for $M_t=172.5$~GeV)}  &  &  &  & \\[3pt] \hline

{\boldmath $|V_{tb}|$}  & & CDF: $|V_{tb}| = 0.91 \pm 0.11 {\rm (stat+sys)} \pm 0.07 {\rm (th)}$ & 1 & 3.2 \\[3pt]
		       & & D0: $|V_{tb}| = 1.02^{+0.10}_{-0.11}$ & & 5.4 \\[3pt] \hline
% {\footnotesize \boldmath $B(t \to Wb)/B(t \to Wq)$} & & CDF: $>0.61$ @ 95\% CL & 1 & 0.2 \\[3pt]
 {\boldmath $\frac{B(t \to Wb)}{B(t \to Wq)}$} & & CDF: $>0.61$ @ 95\% CL & 1 & 0.2 \\[3pt]
		       & & D0: $0.90 \pm 0.04$ & & 5.4 \\[3pt] \hline
% {\footnotesize \boldmath $\sigma(gg \to \ttbar)/\sigma(\ppbar \to \ttbar)$} & $\ppbar \to \ttbar $ & CDF: $0.07^{+0.15}_{-0.07}$ & 0.18 & 1 \\[3pt] \hline
 {\boldmath $\frac{\sigma(gg \to t \bar{t})}{\sigma(p \bar{p} \to t \bar{t})}$}  & $\ppbar \to \ttbar $ & CDF: $0.07^{+0.15}_{-0.07}$ & 0.18 & 1 \\[3pt] \hline
 {\boldmath $M_t$} & & Tev: $173.2 \pm 0.9$~GeV & - & up to 5.8 \\[3pt]
        & & Atlas: $169.3 \pm 6.3$~GeV & - & 0.035 \\[3pt]
        & & CMS:   $173.4 \pm 3.3$~GeV & - & 0.036 \\[3pt] \hline
{\boldmath $M_t - M_{\bar{t}}$} & & CDF: $-3.3 \pm 1.4 {\rm (stat)} \pm 1.0 {\rm (syst)}$ GeV & 0 & 5.6 \\[3pt]
			     & & D0: $0.8 \pm 1.8 {\rm (stat)} \pm 0.5 {\rm (syst)}$ GeV & & 3.6 \\[3pt] \hline
 {\bf W helicity} 	   & & Tev: $f_0 = 0.732 \pm 0.063 {\rm (stat)} \pm 0.052 {\rm (syst)}$ & 0.7 & up to 5.4 \\[3pt] 
 {\bf fraction}		   & & Atlas: $f_0 = 0.59 \pm 0.10 {\rm (stat)} \pm 0.07 {\rm (syst)}$ & 0.7 & 0.035 \\[3pt] \hline
{\bf Charge}		   & & CDF: -4/3 excluded @ 95\% CL & 2/3 & 5.6 \\[3pt]
			   & & D0: 4/3 excluded @ 92\% CL & & 0.37 \\[3pt] \hline
{\boldmath $\Gamma_t$}	   & & CDF: $<7.6$ GeV @ 95\% CL & 1.26 GeV & 4.3 \\[3pt]
			   & & D0: $1.99^{+0.69}_{-0.55}$ GeV & & up to 2.3 \\[3pt] \hline
{\bf spin correlation}      & $C_{\rm beam}$ & CDF: $0.72 \pm 0.64 {\rm (stat)} \pm 0.26 {\rm (syst)}$ & $0.777^{+0.027}_{-0.042}$ & 5.3 \\[3pt]
				   & & D0: $0.57 \pm 0.31 {\rm (stat+sys)}$ & & 5.4  \\[3pt] \hline
 {\bf Charge}    	   & $\ppbar \to \ttbar$ & CDF: $0.158 \pm 0.074$ & 0.06 & 5.3 \\[3pt]
 {\bf asymmetry}	   & & D0: $0.196 \pm 0.065$ & & 5.4 \\[3pt]  
 			   & $pp \to \ttbar$ & Atlas: $A^y_C=-0.024 \pm 0.016 {\rm (stat)} \pm 0.023 {\rm (syst)}$ & 0.006 & 0.7 \\[3pt]
			   & & CMS: $A^\eta_C=-0.016 \pm 0.030 {\rm (stat)} ^{+0.010}_{-0.019} {\rm (syst)}$ & 0.013 & 1.1 \\[3pt]
\hline
\end{tabular}
\end{center}
\caption{Summary of the main top quark properties}
\label{tab:summary}
\end{table}

\end{center}

\end{document}